\documentclass[epj]{svjour}
\usepackage{multirow}
\usepackage{amsmath}
\usepackage[latin1]{inputenc}
\usepackage{array}
\usepackage{color}
\usepackage[caption=false]{subfig}
\usepackage[french]{babel}
\usepackage[pdftex]{graphicx}

\begin{document}
\title{Vacuum magnetic linear birefringence using pulsed fields}
\subtitle{Status of the BMV experiment}
\author{A. Cadène, P. Berceau, M. Fouché, R. Battesti and C. Rizzo
}                     
\institute{Laboratoire National des Champs Magn\'etiques Intenses
(UPR 3228, CNRS-UPS-UJF-INSA), F-31400 Toulouse Cedex, France, EU}
\date{Received: date / Revised version: date}
%
\abstract{We present the current status of the BMV experiment. Our
apparatus is based on an up-to-date resonant optical cavity
coupled to a transverse magnetic field. We detail our data
acquisition and analysis procedure which takes into account the
symmetry properties of the raw data with respect to the
orientation of the magnetic field and the sign of the cavity
birefringence. The measurement result of the vacuum magnetic
linear birefringence $k_\mathrm{CM}$ presented in this paper was
obtained with about 200 magnetic pulses and a maximum field of
6.5\,T, giving a noise floor of about $8 \times
10^{-21}$\,T$^{-2}$ at $3\sigma$ confidence level.
} 
\maketitle
\section{Introduction}
It is known since the beginning of the 20$^\mathrm{th}$ century
that any medium shows a linear birefringence in the presence of a
transverse external magnetic field $\textbf{\textit{B}}$. This
effect is usually known as the Cotton-Mouton (CM) effect (see
Ref.\,\cite{Rizzo1997} and references therein). The existence of
such a magnetic linear birefringence has also been predicted in
vacuum around 1970 in the framework of Quantum ElectroDynamics
(QED) \cite{Bialynicka1970,Ritus1975}. It is one of the non-linear
optical effects described by the Heisenberg-Euler effective
lagrangian (see Ref. \cite{Battesti2013} and references therein)
and it can be seen as the result of the interaction of the
external magnetic field with quantum vacuum fluctuations. In a
vacuum therefore the index of refraction $n_\|$ for light
polarized parallel to $\textbf{\textit{B}}$ is expected to be
different from the index of refraction $n_\bot$ for light
polarized perpendicular to $\textbf{\textit{B}}$ such that
\cite{Battesti2013}:
\begin{eqnarray}
\Delta n_\mathrm{CM} &=& n_\| - n_\bot,\\
&=& k_\mathrm{CM} B^2.
\end{eqnarray}
At the first order in the fine structure constant $\alpha$,
$k_\mathrm{CM}$ can be written as:
\begin{equation}
k_\mathrm{CM} = 2\alpha^2 \hbar^3 / 15 \mu_{0} m_\mathrm{e}^4 c^5,
\end{equation}
with $\hbar$ the Planck constant over $2\pi$, $m_\mathrm{e}$ the
electron mass, $c$ the speed of light in vacuum, and $\mu_0$ the
magnetic constant. Using the CODATA recommended values for
fundamental constants \cite{CODATA}, one obtains:
\begin{equation}
 k_\mathrm{CM}
\sim 4.0 \times10^{-24} \mathrm{T}^{-2}.
\end{equation}

In spite of several experimental attempts, the experimental proof
of such a very fundamental QED prediction is still lacking
\cite{Battesti2013}. All recent experiments, both completed or
running, measure $\Delta n_\mathrm{CM}$ via the ellipticity $\psi$
induced on a linearly polarized light propagating in the
birefringent vacuum:
\begin{equation}\label{Eq:ell}
\psi = \pi k_\mathrm{CM}  \frac{L_B}{\lambda} B^2 \sin
2\theta_\mathrm{P},
\end{equation}
where $\lambda$ is the light wavelength, $L_B$ is the path length
in the magnetic field, and $\theta_\mathrm{P}=45^\circ$ is the
angle between the light polarization and the birefringence axis.
This equation clearly shows that the critical experimental
parameter is the product $B^2L_B$. In order to increase the
ellipticity to be measured, one usually uses an optical cavity to
store light in the magnetic field region as long as possible. The
total acquired ellipticity $\Psi$ is linked to the ellipticity
$\psi$ acquired in the absence of cavity and depends on the cavity
finesse $F$ as:
\begin{eqnarray}
\Psi = \frac{2F}{\pi}\,\psi.
\end{eqnarray}

\begin{figure}[t]
\begin{center}
\includegraphics[width=8cm]{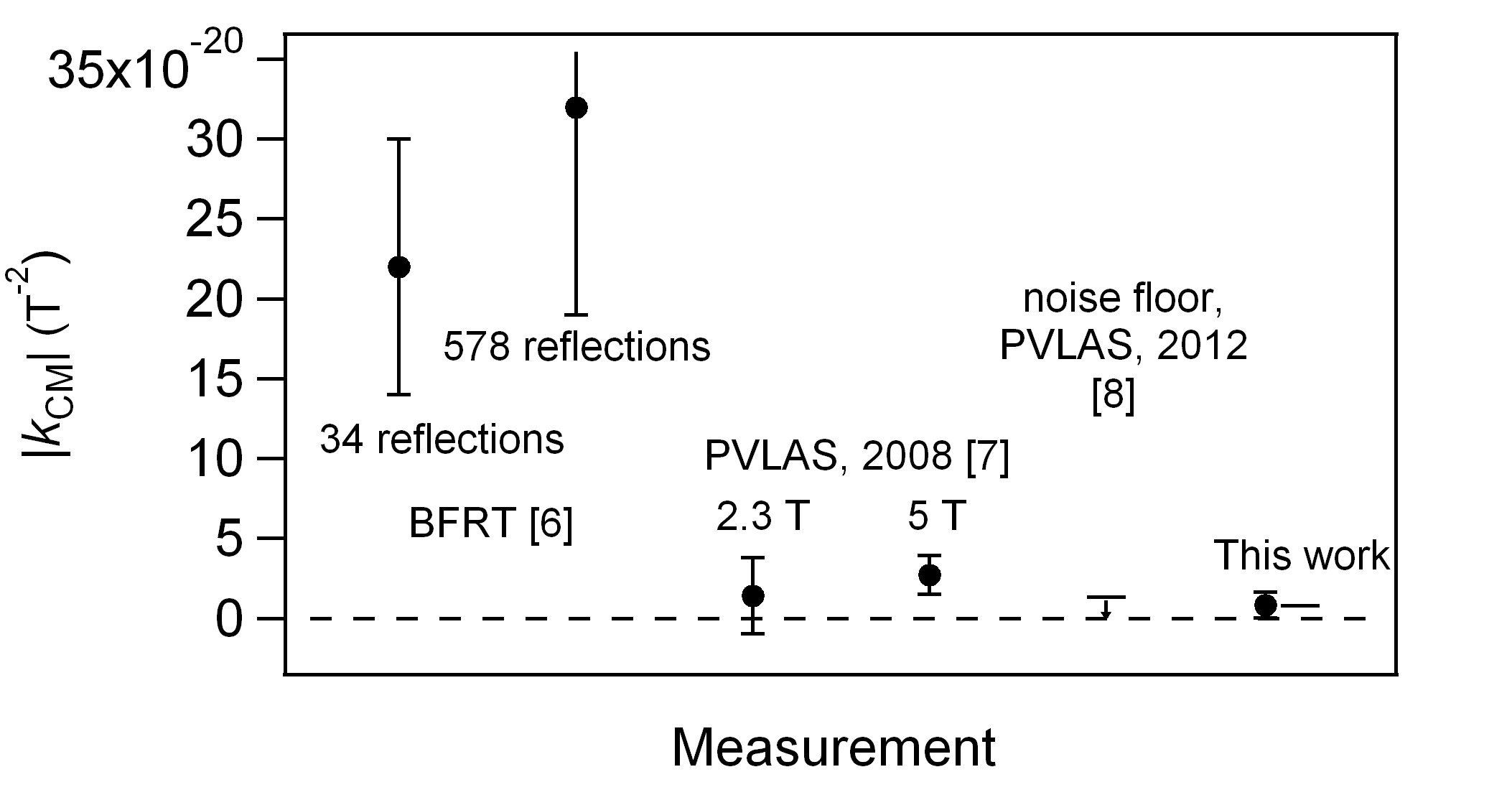}
\caption{\label{Fig:summary_kcm} Comparison of reported absolute
values of the vacuum magnetic linear birefringence and their
uncertainties represented at 3$\sigma$.}
\end{center}
\end{figure}

After the theoretical calculations in the 70s, a first measurement
of the $k_\mathrm{CM}$ value was published by the BFRT
collaboration \cite{Cameron1993}. It was based on a
superconducting magnet providing a maximum field of 3.9 T, and a
multipass optical cavity. Spurious signals were always present
(see Table V(b) in \cite{Cameron1993}). Final results gave
$k_\mathrm{CM} = (2.2 \pm 0.8) \times 10^{-19}$\,T$^{-2}$ at
$3\sigma$ confidence level for 34 refections
inside the cavity, and $k_\mathrm{CM} = (3.2 \pm 1.3) \times
10^{-19}$\,T$^{-2}$ for 578 reflections. In 2008 a new measurement
was published by the PVLAS collaboration using a Fabry-P\'erot
optical cavity and a superconducting magnet providing a 2.3\,T
field: $k_\mathrm{CM} = (1.4 \pm 2.4) \times 10^{-20}$ T$^{-2}$ at
3$\sigma$ \cite{Zavattini2008}. The same experiment at 5\,T gave
$k_\mathrm{CM} = (2.7 \pm 1.2) \times 10^{-20}$\,T$^{-2}$ at
3$\sigma$. More recently a new version of the PVLAS apparatus
based on two 2.5\,T permanent magnets and a Fabry-P\'erot optical
cavity reached a noise floor corresponding to $k_\mathrm{CM} =
1.3\times 10^{-20}$\,T$^{-2}$ at 3$\sigma$, but "only when no
spurious signal was observed" \cite{Zavattini2012}. All over our paper, we give error bars at $3\sigma$ corresponding to a confidence level of 99.8\%, that usually indicates an evidence for a non-zero signal. All these measurements are summarized in Fig.\,\ref{Fig:summary_kcm}. This
clearly shows that vacuum CM measurements are true experimental
challenges and that one has to focus not only on getting the best
optical sensitivity and maximizing the signal to be measured, but
also on minimizing all the unwanted systematic effects by
decoupling the apparatus from their sources and by performing an
appropriate data analysis.

In this paper we present a measurement of $k_\mathrm{CM}$ obtained
using the first generation setup of the BMV
\textit{(Bir\'efrin\-gence Magn\'etique du Vide)} experiment at
the National High Magnetic Field Laboratory of Toulouse, France
\-(LNCMI-T) \cite{EPJD_BMV}. The novelty of this experiment is the
use of pulsed magnetic fields. This method allows to provide the
highest magnetic fields in terrestrial laboratories without
destroying the coil itself \cite{Battesti2013}. Our apparatus is
also based on the use of an infrared Fabry-P\'erot cavity among
the sharpest in the world \cite{Berceau2012}. We calibrated our
experiment using nitrogen gas \cite{Berceau2012}, and recently
published a high precision measurement of the Cotton-Mouton effect
of helium gas compatible with the theoretical prediction
\cite{Cadene2013}. We present our data acquisition and analysis
procedure that takes into account the symmetry properties of the
raw data with respect to the orientation of the magnetic field and
the sign of the cavity birefringence. The measurement result of
the vacuum magnetic linear birefringence $k_\mathrm{CM}$ presented
in this paper was obtained with about 200 magnetic pulses and a
maximum field of 6.5\,T. It corresponds to the best noise floor
ever reached. It is therefore a clear validation of our innovative
experimental method.

\section{Experimental setup}\label{Par:exp_steup}

\subsection{Apparatus}

\begin{figure}[h]
\begin{center}
\includegraphics[width=8cm]{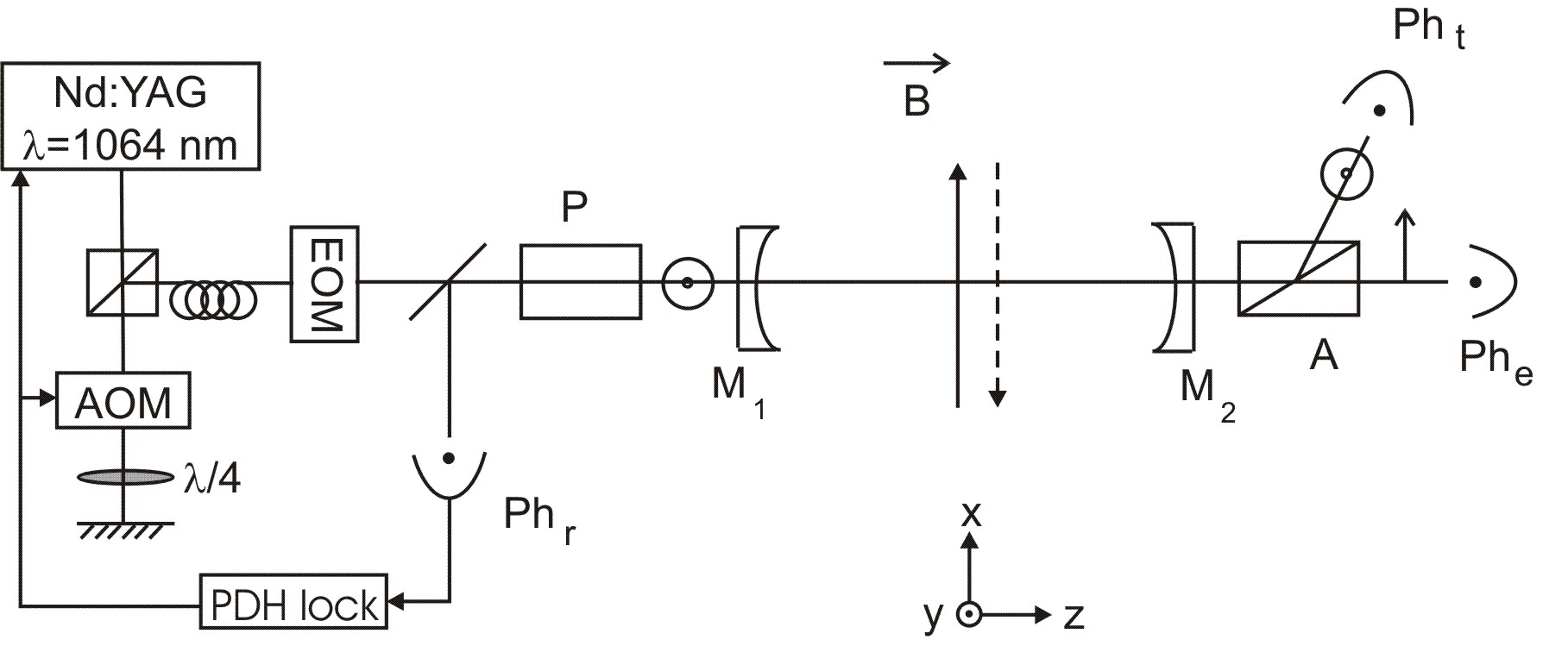}
\caption{\label{Fig:ExpSetup} Experimental setup. EOM,
electro-optic modulator; AOM, acousto-optic modulator; PDH,
Pound-Drever-Hall; Ph, photodiode; P, polarizer; A, analyzer. See
text for more details.}
\end{center}
\end{figure}

Our experimental setup is described in Refs.\,\cite{Cadene2013}.
As shown in Fig.\,\ref{Fig:ExpSetup}, 30 mW of a linearly
polarized Nd:YAG laser beam ($\lambda$ = 1064 nm) goes through an
acousto-optic modulator (AOM) used  in double pass for an
adjustment of the laser frequency. It is then injected into a
monomode optical fiber before entering a high finesse Fabry-Pérot
cavity of length $L_\mathrm{c} = 2.27$\,m, consisting of the
mirrors M$_1$ and M$_2$. This corresponds to a cavity free
spectral range of $\Delta^{\mathrm{FSR}} = c/2 L_\mathrm{c} =
65.996$\,MHz. The laser passes through an electro-optic modulator
(EOM) creating sidebands at 10 MHz. We analyze the beam reflected
by the cavity on the photodiode Ph$_\mathrm{r}$. This signal is
used to lock the laser frequency to the cavity resonance frequency
using the Pound-Drever-Hall method \cite{PDH}, via the
acousto-optic modulator and the piezoelectric and Peltier elements
of the laser.

\begin{figure}[t]
\begin{center}
\includegraphics[width=8cm]{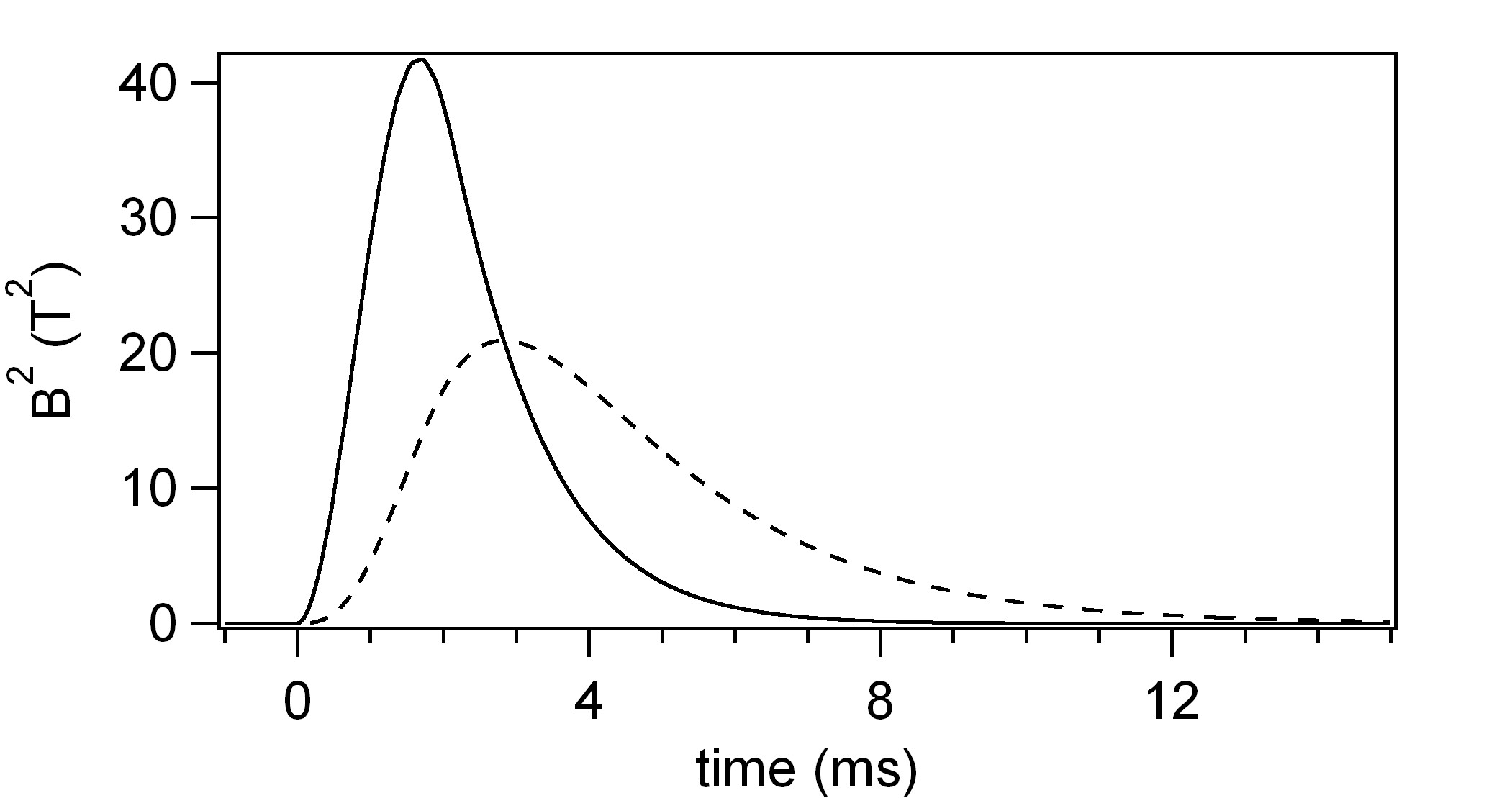}
\caption{\label{Fig:B} Square of the magnetic field amplitude as a
function of time for a maximum field of 6.5\,T. Solid black curve,
$B^2$; dashed curve, $B^2_\mathrm{f}$.}
\end{center}
\end{figure}

To measure the ellipticity induced by the Cotton-Mouton effect one
needs a transverse magnetic field as high as possible. This is
fulfilled using pulsed fields delivered by one magnet, named
X-coil, especially designed in our laboratory. The principle of
this magnet and its properties are described in details in
Refs.\,\cite{EPJD_BMV,Batut2008}. It can provide a maximum field
of more than 14\,T over an equivalent length $L_B$ of 0.137\,m
\cite{Berceau2012}. Data have been taken with a maximum magnetic
field of 6.5\,T reached within 1.70\,ms while the total duration
of a pulse is less than 10\,ms as shown in Fig.\,\ref{Fig:B}.
Moreover, we can remotely switch the high-voltage connections to
reverse $\textit{\textbf{B}}$ in order to set it parallel or
antiparallel to the \textit{x} direction. The maximum repetition
rate is 6 pulses per hour.

We infer the cavity finesse from the measurement of the photon
lifetime $\tau$ \cite{Berceau2012}. Its value is regularly checked
during data taking and we get $\tau = 1.07$\,ms. The corresponding
finesse is:
\begin{equation}
F = \frac{\pi c \tau}{L_\mathrm{c}},
\end{equation}
We get $F = 445\,000$ with a relative variation that does not
exceed 6\,$\%$ at the 3$\sigma$ confidence level. This corresponds
to a cavity linewidth $\Delta \nu = c/2FL_\mathrm{c}$ of 148 Hz.
This is one of the sharpest infrared cavity in the world
\cite{Berceau2012}.

Before entering the Fabry-P\'erot cavity, light is polarized by
the polarizer P. The beam transmitted by the cavity is then
analyzed by the analyzer A crossed at maximum extinction. We
extract both polarizations: parallel and perpendicular to P. The
extraordinary ray, whose polarization is perpendicular to the
incident polarization, is detected by the photodiode
Ph$_\mathrm{e}$ (power $I_\mathrm{e}$), while the ordinary ray,
whose polarization is parallel to the incident polarization, is
detected by Ph$_\mathrm{t}$ (power $I_\mathrm{t}$).

All the optical devices from the polarizer to the analyzer are
placed in an ultrahigh-vacuum chamber. During operation, the
pressure inside the UHV vessel was about $10^{-7}$\,mbar. We have
monitored the vacuum quality with a residual gas analyzer.
Residual gases can cause a measurable CM effect. Most important
contributions come from N$_2$ and O$_2$ leading to a
$k_\mathrm{CM}$ of $1.5\times10^{-23}$\,T$^{-2}$. Moreover
dielectric mirrors also induce a CM effect corresponding to an
ellipticity of $8\times 10^{-10}$\,rad.T$^{-2}$ per reflection, as
reported in Ref.\,\cite{CM_Miroir}. The stray transverse magnetic
field at the mirror position is smaller than 150\,$\mu$T, giving
in our case $k_\mathrm{CM} < 1\times10^{-24}$\,T$^{-2}$. We expect
these CM effects to be smaller than the measured noise floor.

\subsection{Signals}\label{Sec:Signals}

The ellipticity $\Psi(t)$ induced by the transverse magnetic field
is related to the ratio of the extraordinary and ordinary powers
as follows:
\begin{eqnarray}\label{Eq:Ellipticitymeas}
\frac{I_\mathrm{e}(t)}{I_\mathrm{t, f}(t)} &=& \sigma^2 +
[\Gamma + \Psi(t)]^2,\nonumber\\
&\simeq& \sigma^2 + \Gamma^2 + 2\Gamma\Psi(t) \;\;\mathrm{for}\;\;
\Psi \ll \Gamma,
\end{eqnarray}
with $\sigma^2$ the polarizer extinction ratio and $\Gamma$ the
total static ellipticity. This static ellipticity is due to the
mirrors' intrinsic phase retardation \cite{BirMirror}. Each mirror
can be regarded as a wave plate. The combination of both wave
plates gives a single wave plate with a total phase retardation
and an axis orientation that depend on each mirror phase
retardation and on their relative orientation \cite{Jacob,brandi}.
Thus, we adjust the value of $\Gamma$ by rotating the mirrors
M$_1$ and M$_2$ around the $z$ axis corresponding to the axis of
light propagation.

To measure the polarizer extinction ratio, we first set $\Gamma =
0$, with no magnetic field. We get $I_\mathrm{e}/I_\mathrm{t,f} =
\sigma^2 \sim 7\times 10^{-7}$. Then, to reach the best
sensitivity, we need $\Gamma^2 \sim \sigma^2$ \cite{EPJD_BMV}.
Starting from $\Gamma = 0$ and rotating M$_1$ in the clockwise or
counterclockwise direction, we choose the value of $\Gamma$ as
well as its sign determined by CM measurements in nitrogen and
helium gas. The measurement of $\sigma^2$ and the adjustment of
the value and sign of $\Gamma$ are done before each magnetic
pulse.

Due to the photon lifetime, the cavity acts as a first order low
pass filter, as explained in details in Ref.\,\cite{Berceau2010}.
Its complex response function $H(\nu)$ is given by:
\begin{eqnarray}
H(\nu)=\frac{1}{1+i\frac{\nu}{\nu_{\mathrm{c}}}},\label{Eq:H_nu}
\end{eqnarray}
with $\nu$ the frequency and $\nu_{\mathrm{c}} = 1/4\pi\tau \simeq
74$\,Hz the cavity cutoff frequency. This filtering has to be
taken into account in particular for the time dependent magnetic
field applied inside the Fabry-Pérot cavity. The ellipticity
$\Psi$ induced by the external magnetic field is thus proportional
to $B^2_\mathrm{f}$:
\begin{equation}
\Psi(t) = \alpha B^2_\mathrm{f}(t),\label{Eq:Psi_Bf}
\end{equation}
where the filtered field $B^2_\mathrm{f}$ is calculated from $B^2$
taking into account the cavity filtering. The time profile of
$B^2_\mathrm{f}$ is plotted in Fig.\,\ref{Fig:B} with the dashed
curve. In particular, the cavity filtering induces an attenuation
and a shift of the maximum. The cavity filtering has also to be
applied to $I_\mathrm{t}$ as explained in details in
Refs.\,\cite{Berceau2010,Cadene2013}.

The calculated signals used for the analysis are described in
details in Ref.\,\cite{Cadene2013}. In order to extract the
ellipticity $\Psi(t)$ from Eq.\,(\ref{Eq:Ellipticitymeas}), we
calculate the following $Y(t)$ signal after each pulse:
\begin{eqnarray}
Y(t) &=& \frac{\frac{I_\mathrm{e}(t)}{I_\mathrm{t,f}(t)}-
I_{\mathrm{dc}}} {2\mid\Gamma\mid}, \label{Eq:IeIt-DC}\\
&\simeq& \gamma \Psi(t), \label{Eq:Y_gammaPsi}
\end{eqnarray}
where $\gamma$ corresponds to the sign of $\Gamma$. We calculate
the static signal $I_{\mathrm{dc}} = \sigma^2 + \Gamma^2$ before
the pulse as follows:
\begin{equation}
I_{\mathrm{dc}} = \bigg<
\frac{I_{\mathrm{e}}(t)}{I_{\mathrm{t,f}}(t)}\bigg>\bigg|_{t_\Gamma<t<0},
\end{equation}
where $t_\Gamma$ corresponds to the beginning of the analysis and
$t = 0$ to the beginning of the applied magnetic field. The
absolute value of the cavity ellipticity is measured a few
milliseconds before each magnetic pulse thanks to the following
equation:
\begin{eqnarray}\label{Eq:Gamma}
|\Gamma| = \sqrt{\bigg<
\frac{I_{\mathrm{e}}(t)}{I_{\mathrm{t,f}}(t)}\bigg>\bigg|_{t_\Gamma<t<0}
- \sigma^2}.
\end{eqnarray}

Signals $Y(t)$ are collected for both signs of $\Gamma$ and for
both directions of $\textit{\textbf{B}}$: parallel to $x$ is
denoted as $>0$ and antiparallel is denoted as $<0$. This gives
four data series: ($\Gamma>0$, $B>0$), ($\Gamma>0$, $B<0$),
($\Gamma<0$, $B<0$) and ($\Gamma<0$, $B>0$). For each series,
signals calculated with Eq.\,(\ref{Eq:IeIt-DC}) are averaged and
denoted as $Y_{>>}$, $Y_{><}$, $Y_{<<}$ and $Y_{<>}$. The first
subscript corresponds to $\Gamma >0$ or $<0$ and the second one
corresponds to $\textit{\textbf{B}}$ parallel or antiparallel to
$x$.

\section{Data analysis and results}

\begin{figure}[t]
\begin{center}
\includegraphics[width=8cm]{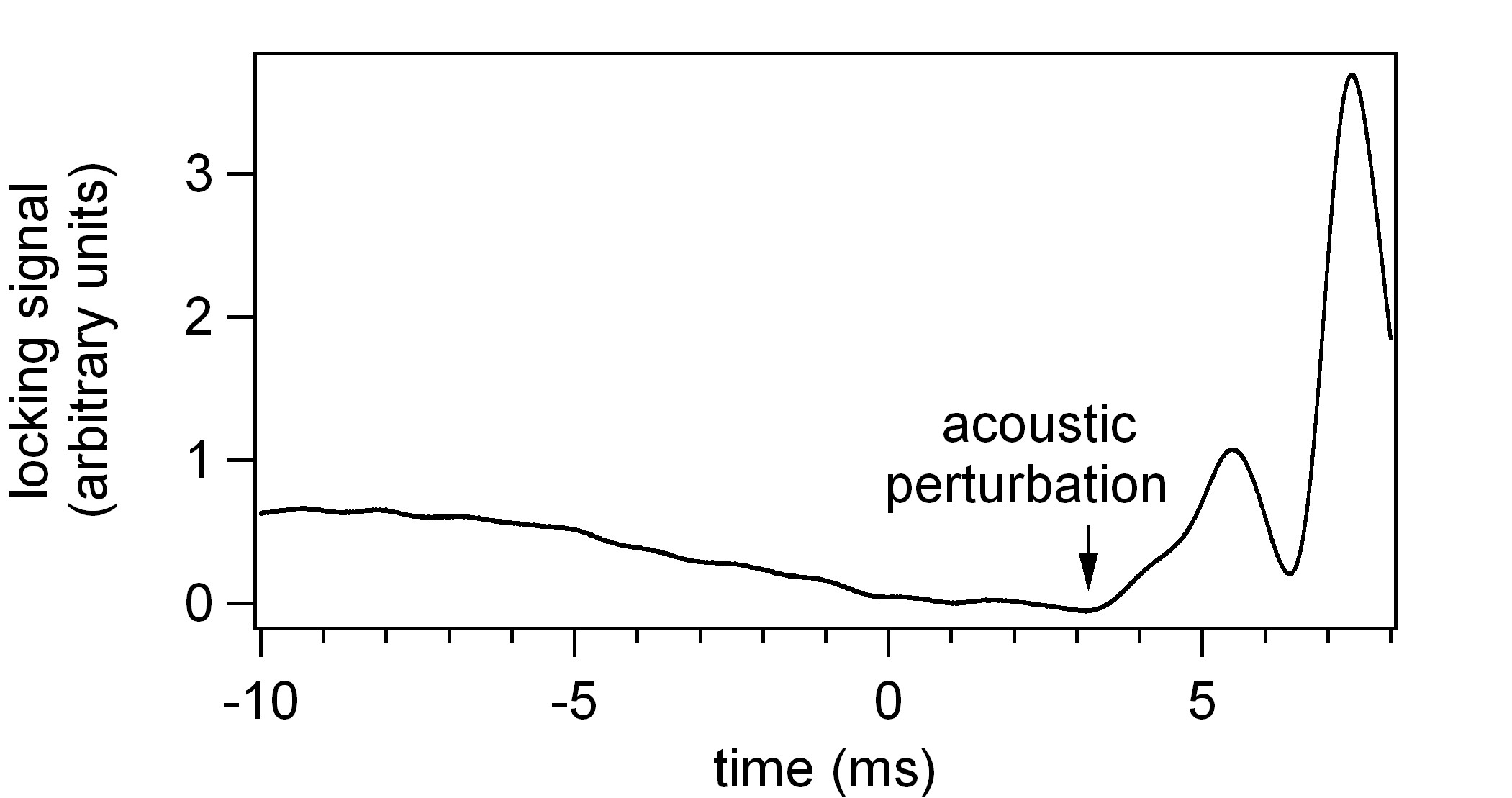}
\caption{\label{Fig:locking_signal} Time evolution of the locking
signal during a magnetic pulse. The magnetic field is applied at
$t=0$\,ms.}
\end{center}
\end{figure}

The raw signals, such as $I_\mathrm{t}(t)$, $I_\mathrm{e}(t)$,
$B(t)$ or the cavity locking signal, are recorded 25\,ms before
the beginning of the magnetic field and 25\,ms after. A typical
cavity locking signal is plotted in
Fig.\,\ref{Fig:locking_signal}. We clearly see a perturbation
which begins at about 3.2\,ms. This corresponds to the acoustic
perturbation triggered at $t = 0$ by the magnetic pulse. This
perturbation travels trough the air to the mirror mounts. We have
confirmed the arrival time on the mirror mounts with
accelerometers. This perturbation induces an ellipticity noise
which degrades our sensitivity. We have thus decided to stop the
analysis at $t = 3.1$\,ms. Symmetrically, we start the analysis at
$t_\Gamma = -3.1$\,ms. It also allows to avoid drifts and long
time variations of $\Gamma$.

\begin{center}
\begin{figure}[htp!!]
\centering
\subfloat[]{\includegraphics[width=4.2cm]{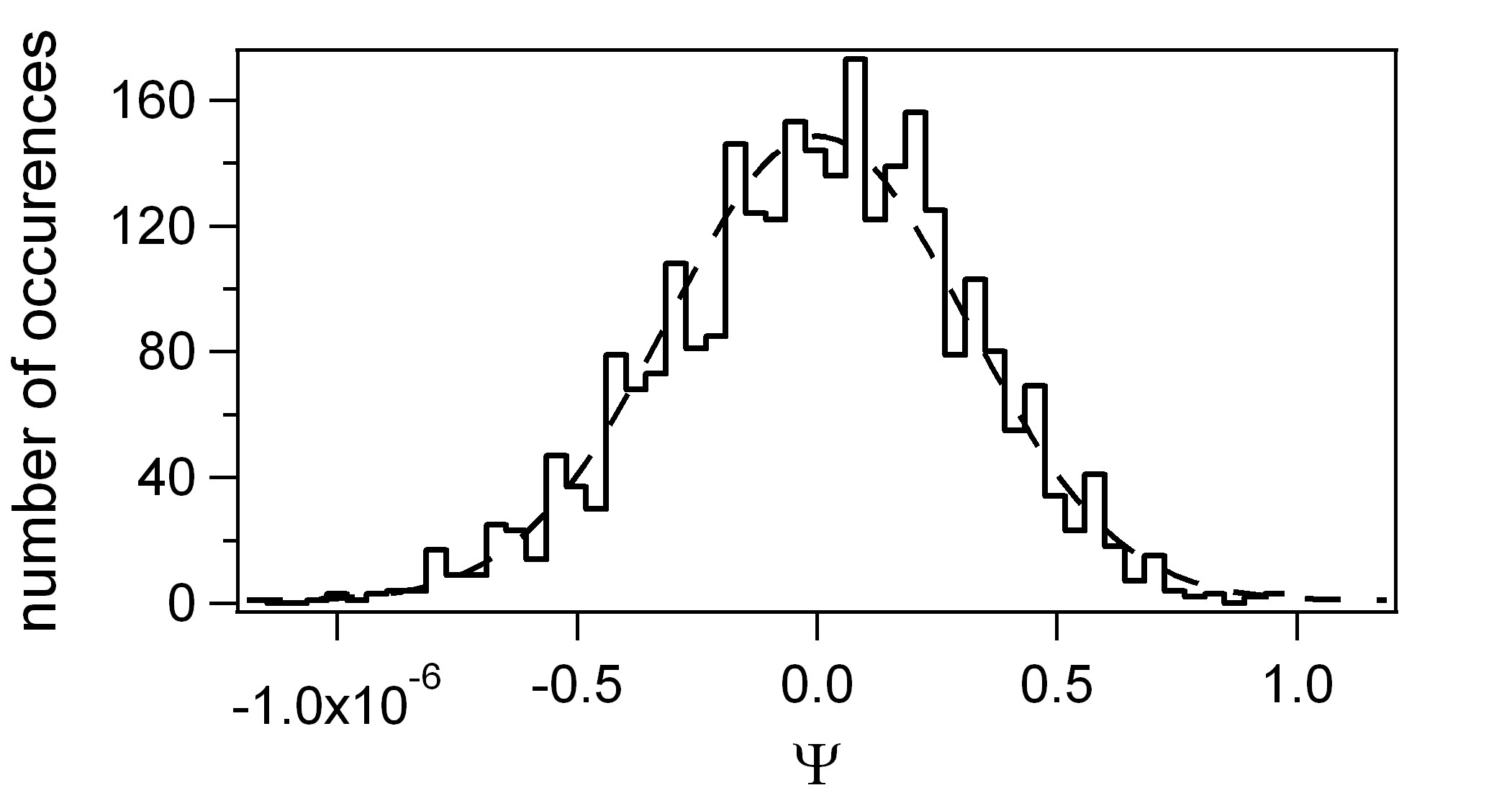}\label{Fig:Histogram_Good}}
\subfloat[]{\includegraphics[width=4.2cm]{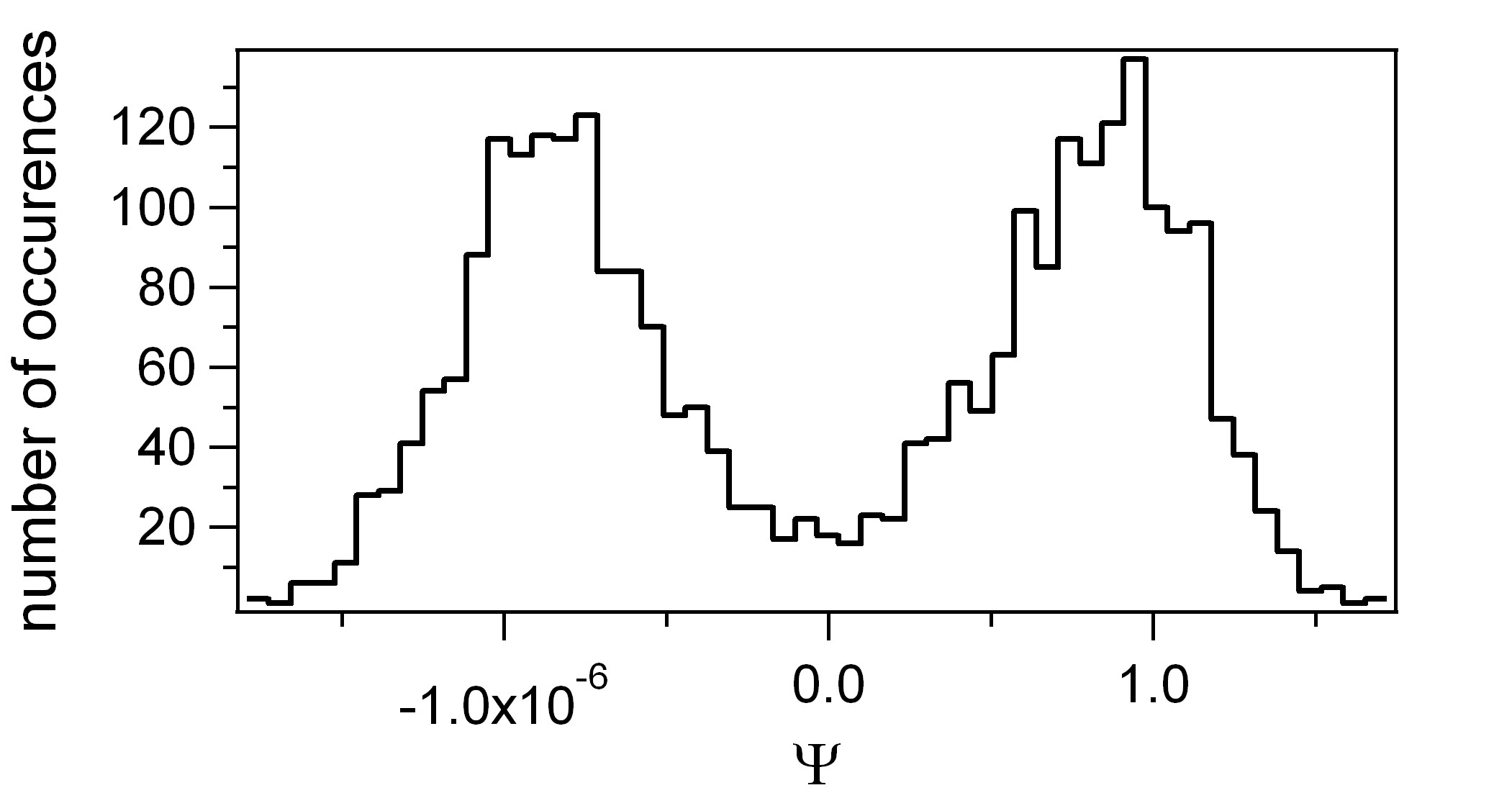}\label{Fig:Histogram_Bad}}
\caption{\small{Typical histogram of $\Psi(t)$ before the magnetic
pulse. a) The histogram can be fitted by a gaussian function
(dashed curve): the shot is selected. b) Rejected shot.}}
\end{figure}
\end{center}

For each pulse applied in vacuum, we first calculate the
$|\Gamma|$ value following Eq.\,(\ref{Eq:Gamma}). To check that
this corresponds to a meaningful value, we plot the histogram of
the following signal for $t_\Gamma<t<0$:
\begin{equation}
\Psi(t) =
\sqrt{\frac{I_\mathrm{e}(t)}{I_\mathrm{t,f}(t)}-\sigma^2} -
\Gamma.
\end{equation}
This corresponds to 3100 values acquired every 1\,$\mu$s. With
white noise and because no induced ellipticity is present at
$t<0$, the histogram is centered on 0 and corresponds to a
gaussian distribution, as shown in Fig.\,\ref{Fig:Histogram_Good}.


\begin{figure}[htp!!]
\begin{center}
\includegraphics[width=8cm]{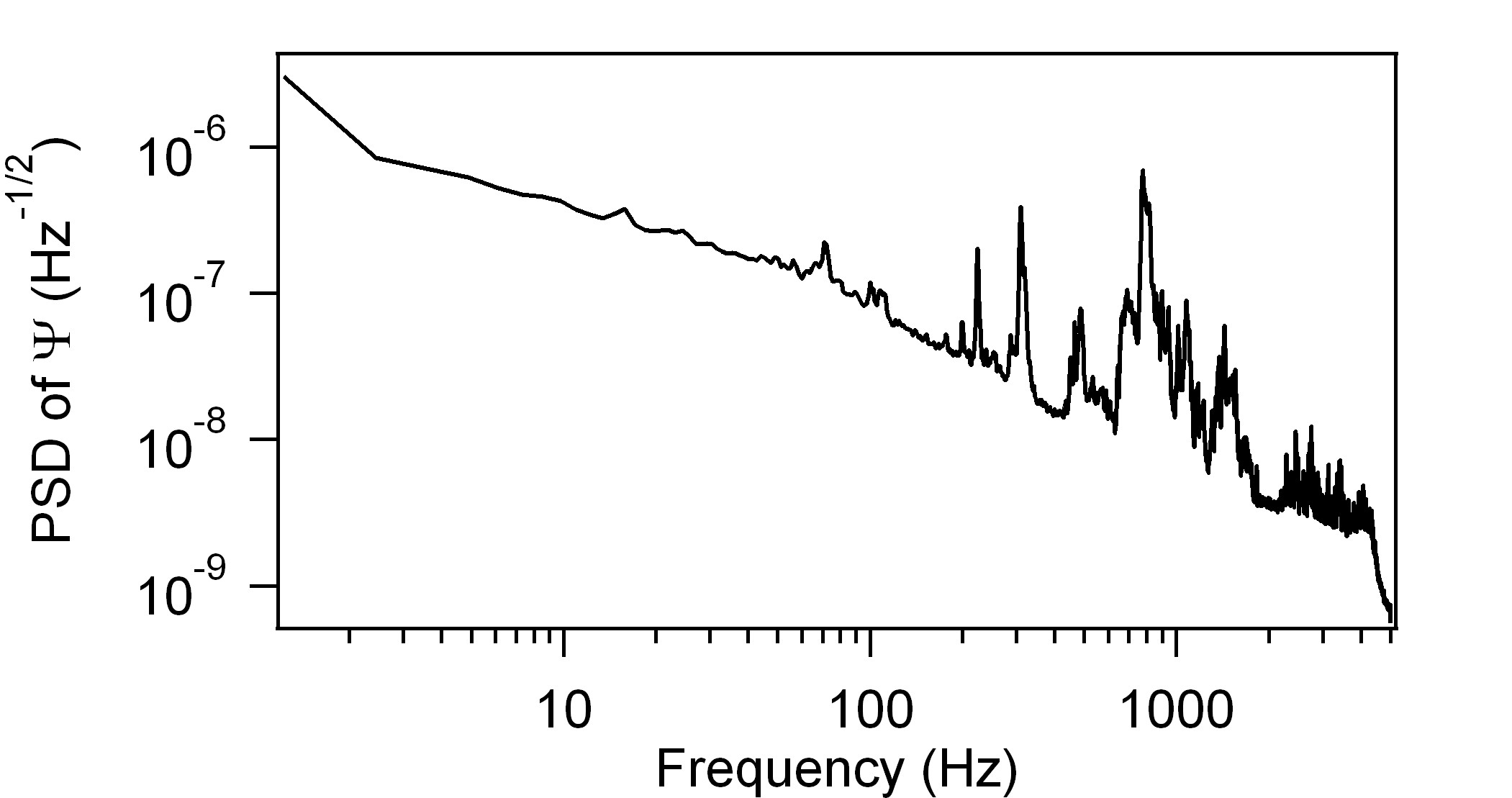}
\caption{\label{Fig:DSP_Psi} Power spectral density of $\Psi$ in
the absence of the magnetic field. We observe the different
mechanical resonances of the setup.}
\end{center}
\end{figure}

However, some of the histograms cannot be fitted by a gaussian
function, as shown in Fig.\,\ref{Fig:Histogram_Bad}. The main
origin of this type of distributions is mechanical oscillations of
the setup induced by the environment and leading to static
ellipticity fluctuations, event if the magnetic filed is not
applied. These mechanical oscillations can be directly observed on
the power spectral density (PSD) of the ellipticity $\Psi$ in the
absence of the magnetic field, as shown in
Fig.\,\ref{Fig:DSP_Psi}. In the case corresponding to
Fig.\,\ref{Fig:Histogram_Bad}, we cannot give a statistical and
significant value of $\Gamma$. The corresponding shots are thus
rejected. Finally we selected 101 pulses. It should be noted that
this selection is performed for $t<0$, thus before the magnetic
pulse. We do not select or reject pulses with an analysis on the
signal we want to measure, thus induced by the magnetic field at
$t>0$.

From the 101 selected pulses, we calculate the signals $Y_{>>}$,
$Y_{><}$, $Y_{<<}$, and $Y_{<>}$, denoted by $Y_\mathrm{j}$ with
$\mathrm{j} = >>,\,><,\,>>,\,<>$. As explained in
section\,\ref{Sec:Signals}, they correspond to the average of the
$Y(t)$ signals calculated with Eq.\,(\ref{Eq:IeIt-DC}) for each of
the four series. The $Y_\mathrm{j}$ uncertainties are calculated
at each time $t_\mathrm{i}$, $\Delta
Y_\mathrm{j}\mathrm(t_\mathrm{i}) =
\sigma_\mathrm{j}\mathrm(t_\mathrm{i})/\sqrt{N_\mathrm{j}}$, with
$\sigma_\mathrm{j}\mathrm(t_\mathrm{i})$ the standard deviation of
the $Y_\mathrm{j}\mathrm(t_\mathrm{i})$ distribution and
$N_\mathrm{j}$ the number of shots for the $\mathrm{j}$ series.

As explained in details in Ref.\,\cite{Cadene2013}, one
has to consider systematic effects that mimic the CM effect we
want to measure. We thus analyze our data following a general expression taking
into account the symmetry properties of $Y_\mathrm{j}$ towards
experimental parameters:
\begin{eqnarray*}\label{Eq:Y}
Y_{>>} &=& a_{>>}S_{++}+b_{>>}S_{+-}+c_{>>}S_{--}+d_{>>}S_{-+},\\
&=& a_{>>}S_{++}+b_{>>}S_{+-}+c_{>>}S_{--}+\Psi,\\
Y_{><} &=& a_{><}S_{++}-b_{><}S_{+-}-c_{><}S_{--}+d_{><}S_{-+},\\
&=& a_{><}S_{++}-b_{><}S_{+-}-c_{><}S_{--}+\Psi,\\
Y_{<<} &=& a_{<<}S_{++}-b_{<<}S_{+-}+c_{<<}S_{--}-d_{<<}S_{-+},\\
&=& a_{<<}S_{++}-b_{<<}S_{+-}+c_{<<}S_{--}-\Psi,\\
Y_{<>} &=& a_{<>}S_{++}+b_{<>}S_{+-}-c_{<>}S_{--}-d_{<>}S_{-+},\\
&=& a_{<>}S_{++}+b_{<>}S_{+-}-c_{<>}S_{--}-\Psi.
\end{eqnarray*}
The $S$ functions correspond to a given symmetry towards the sign
of $\Gamma$ and the direction of $\textit{\textbf{B}}$. The first
subscript $+$ (resp. $-$) indicates an even (resp. odd) parity
with respect to the sign of $\Gamma$. The same convention is used
for the second subscript corresponding to $\textit{\textbf{B}}$.
Each $S$ function has a different physical origin which are
summarized in Tab.\,\ref{Tab:S_effects}. CM effect signal
contributes to $S_{-+}$ since it depends on the cavity
birefringence $\Gamma$ and on the square of the magnetic field
amplitude as shown in Eqs.\,(\ref{Eq:ell}) and
(\ref{Eq:Y_gammaPsi}). We can thus replace $d S_{-+}$ by $\gamma
\Psi$.

\begin{center}
\begin{table}[h]
\begin{center}
\begin{tabular}{m{2cm} m{4cm}}
\hline \hline \centering $S$ signal &  \centering Physical effect
\tabularnewline \hline \centering $S_{++}(t)$ & \centering
$\Theta_{\mathrm{F}}^2(t)$, $\Psi^2(t)$ \tabularnewline \centering
$S_{+-}(t)$ & \centering \textit{\textbf{B}} effects on
photodiodes \tabularnewline \centering $S_{--}(t)$ & \centering
$\gamma\Theta_{\mathrm{F}}(t)$ \tabularnewline \centering
$S_{-+}(t)$ & \centering $\gamma\Psi(t)$\tabularnewline \hline
\hline
\end{tabular}
\end{center}
\caption{Possible physical effects contributing to the $S$
signals. The $\Theta_\mathrm{F}$ signal corresponds to a
polarization rotation angle due to the circular birefringence
induced by a longitudinal magnetic field (Faraday effect).}
\label{Tab:S_effects}
\end{table}
\end{center}

\begin{figure}[t]
\begin{center}
\includegraphics[width=8cm]{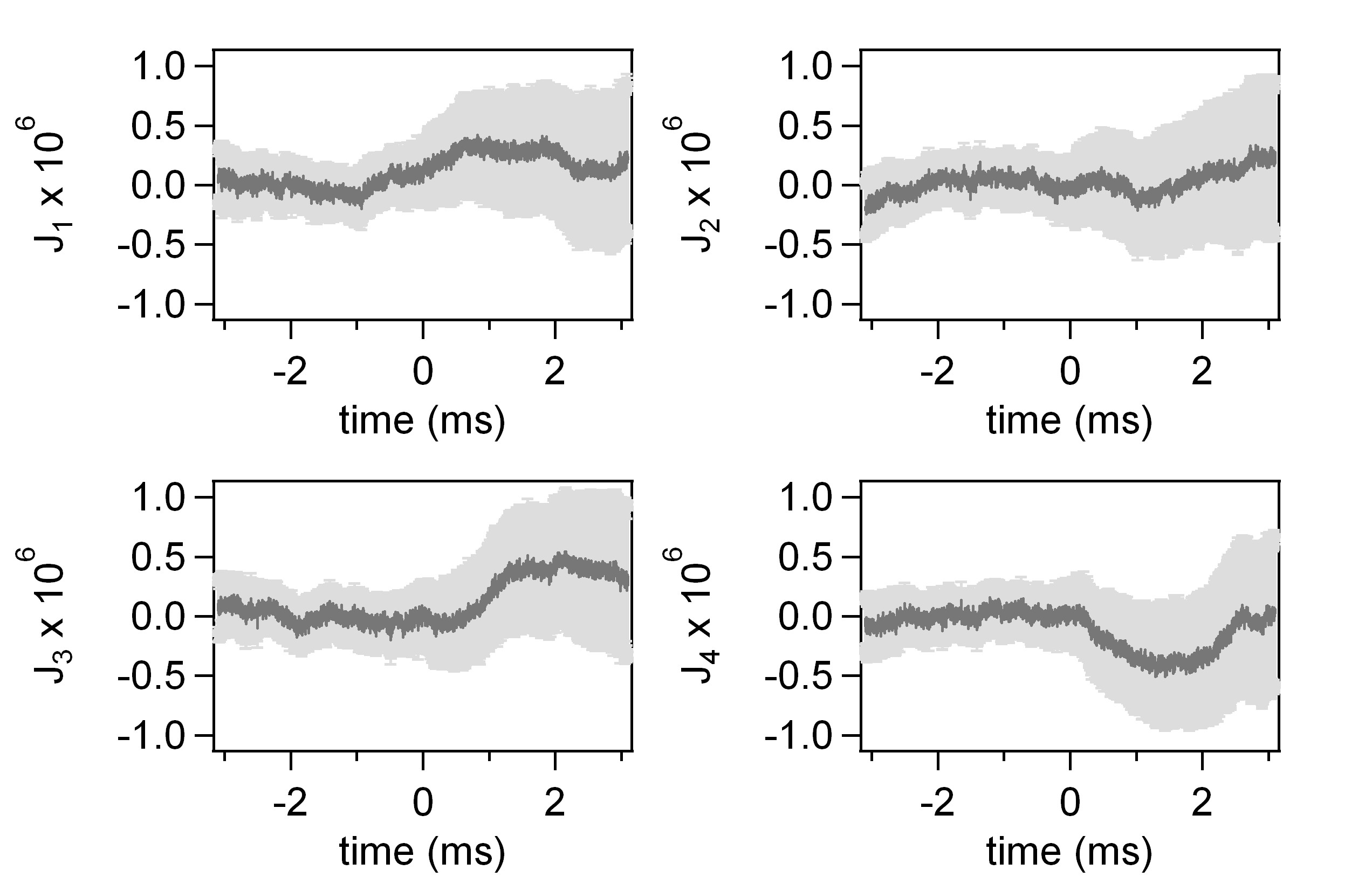}
\caption{\label{Fig:J1234} Time evolution of $J_1$, $J_2$, $J_3$
and $J_4$ (dark grey curve) and their uncertainties at 3$\sigma$
confidence level (light grey).}
\end{center}
\end{figure}

The $S$ functions are then extracted with a linear combination of
$Y_\mathrm{j}$ as follows:
\begin{eqnarray}
J_1 &\equiv& \frac{Y_{>>}+Y_{><}+Y_{<<}+Y_{<>}}{4},\nonumber\\
&=& \overline{a}~S_{++} + \Delta b_1~S_{+-} + \Delta c_1~S_{--} + \Delta d_1~S_{-+},\nonumber\\
J_2 &\equiv& \frac{Y_{>>}-Y_{><}-Y_{<<}+Y_{<>}}{4},\nonumber\\
\nonumber&=& \Delta a_2~S_{++} + \overline{b}~S_{+-}+ \Delta c_2~S_{--} + \Delta d_2~S_{-+},\nonumber\\
J_3 &\equiv& \frac{Y_{>>}-Y_{><}+Y_{<<}-Y_{<>}}{4},\nonumber\\
\nonumber&=& \Delta a_3~S_{++} + \Delta b_3~S_{+-} + \overline{c}~S_{--}+ \Delta d_3~S_{-+},\nonumber\\
J_4 &\equiv& \frac{Y_{>>}+Y_{><}-Y_{<<}-Y_{<>}}{4},\\
\nonumber &=& \Delta a_4~S_{++} + \Delta b_4~S_{+-} + \Delta
c_4~S_{--} +\overline{d}~S_{-+}. \label{Eq:J1234}
\end{eqnarray}
$J_1(t)$, $J_2(t)$, $J_3(t)$ and $J_4(t)$ are plotted in
Fig.\,\ref{Fig:J1234}. Their uncertainties are calculated from the
$Y_\mathrm{j}$ uncertainties. The weighting parameters $a$, $b$,
$c$ and $d$ depend on the experimental adjustment from pulse to
pulse and from day to day. Their relative variations are small:
$\Delta a/\overline{a}$, $\Delta b/\overline{b}$, $\Delta
c/\overline{c}$, $\Delta d/\overline{d}\ll 1$. $\Delta a$, $\Delta
b$ and $\Delta c$ are mainly due to the $\Gamma$ variation from
one shot to another and we can precisely calculate them since
$\Gamma$ is measured for each shot. We obtain $\Delta
a_4/\overline{a} = 5.97 \times 10^{-2}$, $\Delta b_4/\overline{b}
= -7.67 \times 10^{-2}$ and $\Delta c_4/\overline{c} = -8.27
\times 10^{-2}$. These values are of the same order of magnitude
as the one obtained during the CM measurement of helium
gaz\,\cite{Cadene2013}. $\Delta d$ is independent of the $\Gamma$
variation. It mainly comes from a variation of the magnetic field
from one pulse to another. As the $B$ relative variation is small
compared to the $\Gamma$ relative variation we consider $\Delta d
\simeq 0$. The variation of $\Psi$ is thus neglected.

We thus write:
\begin{eqnarray}
J_1 &\simeq& \overline{a}~S_{++},\nonumber\\
J_2 &\simeq& \overline{b}~S_{+-},\nonumber\\
J_3 &\simeq& \overline{c}~S_{--},\nonumber\\
J_4 &\simeq& \frac{\Delta a_4}{\overline{a}} J_1 + \frac{\Delta
b_4}{\overline{b}} J_2 + \frac{\Delta c_4}{\overline{c}} J_3 +
\Psi. \label{Eq:J1234_simple}
\end{eqnarray}

\begin{figure}[t]
\begin{center}
\includegraphics[width=8cm]{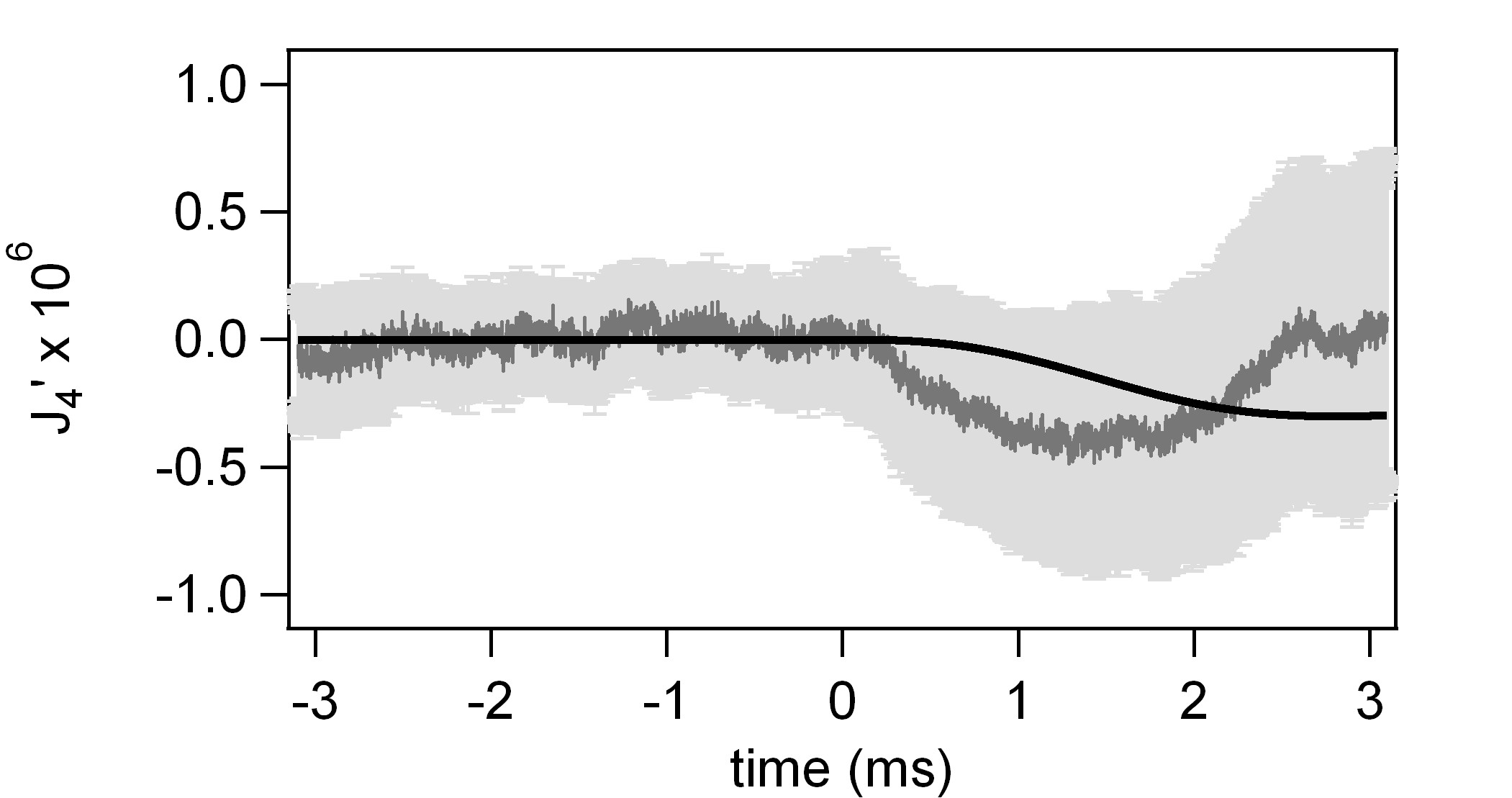}
\caption{\label{Fig:J4p_fit_B2f} Dark grey curve: time evolution
of $J_4'$ and its 3$\sigma$ uncertainties (light grey). Black
curve: $\alpha B^2_\mathrm{f}$ function superimposed to guide the eyes.}
\end{center}
\end{figure}

We then calculate:
\begin{eqnarray}
J_4' &\equiv& J_4 - \bigg[\frac{\Delta a_4}{\overline{a}} J_1 +
\frac{\Delta
b_4}{\overline{b}} J_2 + \frac{\Delta c_4}{\overline{c}} J_3\bigg],\nonumber\\
 &\simeq& \Psi,
\end{eqnarray}
which corresponds to the Cotton-Mouton signal. It is plotted in
Fig.\,\ref{Fig:J4p_fit_B2f} together with a $\alpha B^2_\mathrm{f}$ function superimposed to guide the eyes. Nevertheless, we see that the major component of $J_4'$ is not $\alpha B^2_\mathrm{f}$ but a
supplementary systematic effect.

\begin{center}
\begin{figure}[htp!!]
\centering
\subfloat[Time evolution of $J_4'$. Black curve: fit with a sine function at 180\,Hz.]{\label{Fig:J4p_fit_sin}\includegraphics[width=8cm]{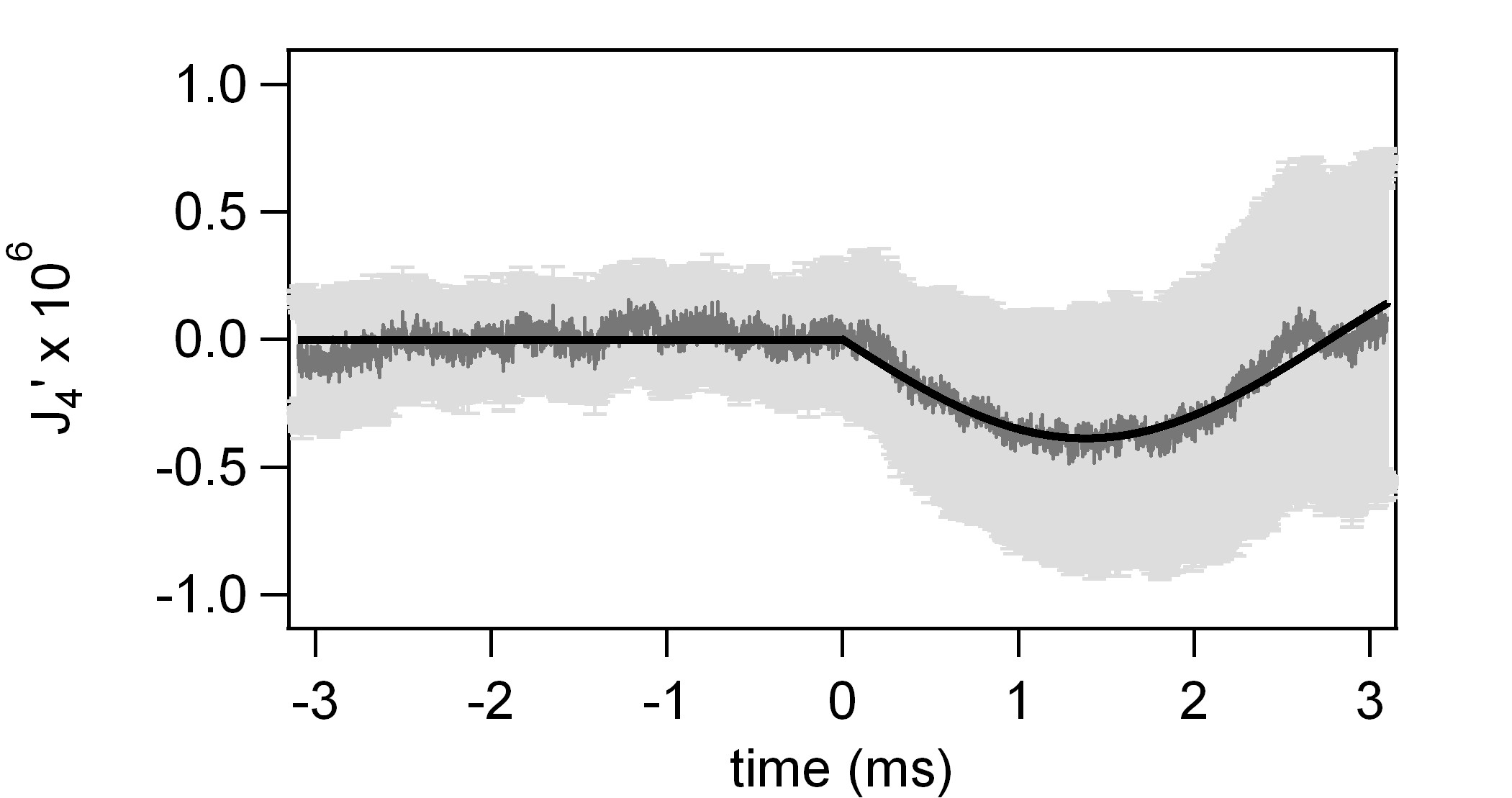}}\\
\subfloat[Time evolution of the residues of $J_4'$. Black curve:
fit with $\alpha
B^2_\mathrm{f}$.]{\label{Fig:J4p_fit_sin_180_B2f}\includegraphics[width=8cm]{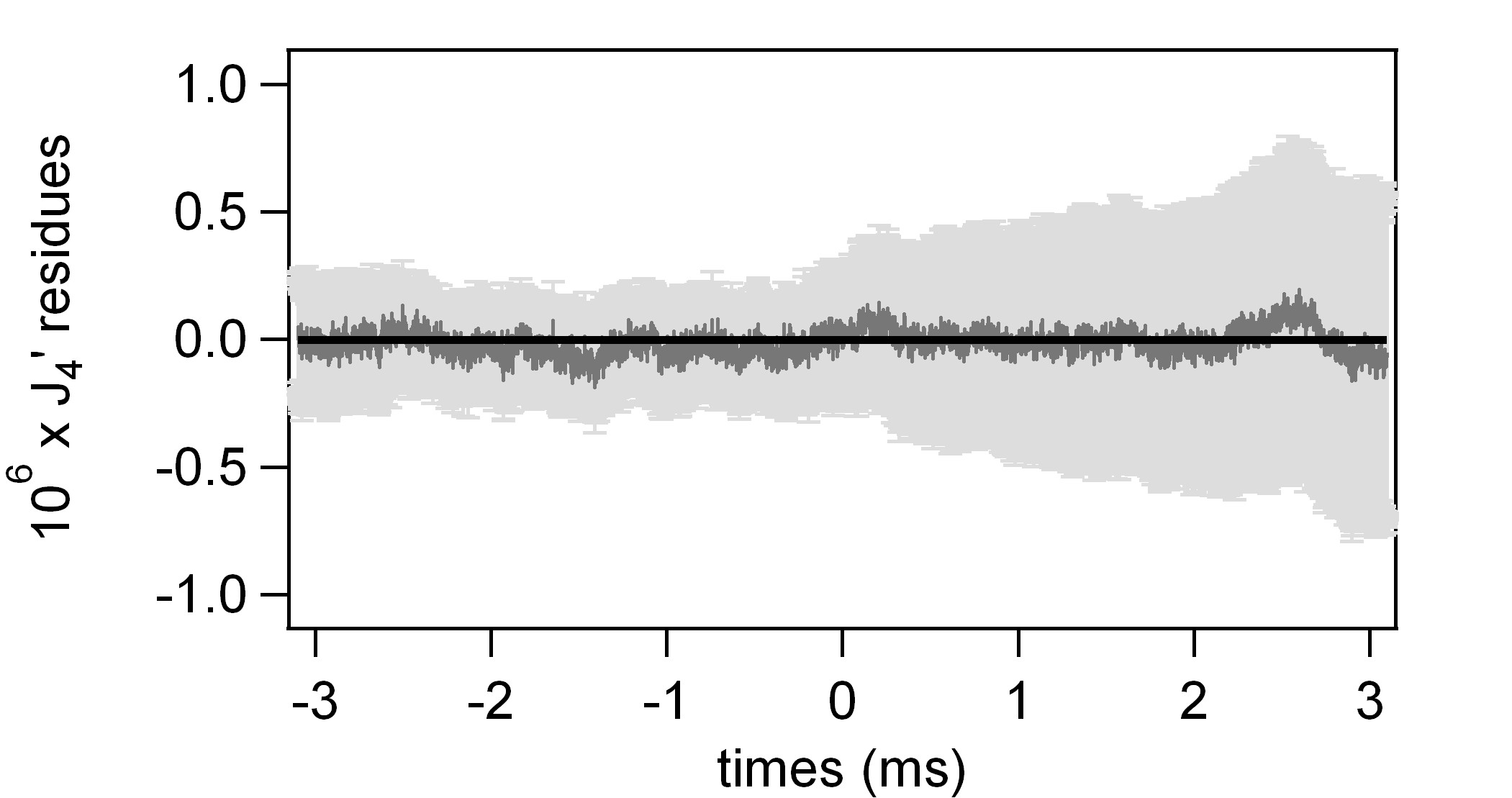}}
\caption{\small{Time evolution of $J_4'$ and its residues (dark
grey). The 3$\sigma$ uncertainties are superimposed in light
grey.}}
\end{figure}
\end{center}

As sais before, the setup is subject to several mechanical
resonances which can be excited both by the environment and the
magnetic field. The latter could thus trigger a mechanical
oscillation of the setup at $t=0$. We try to fit $J_4'$ by a sine
function starting at $t=0$. The fit gives a frequency of ($180 \pm
3$)\,Hz and it is superimposed to $J_4'$ in
Fig.\,\ref{Fig:J4p_fit_sin}. We finally fit the residues by
$\alpha B^2_\mathrm{f}$. The fit is superimposed to the residues
of $J_4'$ in Fig.\,\ref{Fig:J4p_fit_sin_180_B2f}. The
Cotton-Mouton constant $k_\mathrm{CM}$ is deduced from the
measured experimental parameters as follows \cite{Berceau2012}:
\begin{equation}
k_\mathrm{CM} = \frac{\alpha}{4\pi\tau
\Delta^\mathrm{FSR}}\frac{\lambda}{L_B}\frac{1}{\sin
2\theta_\mathrm{P}}.
\end{equation}
We obtain:
\begin{equation}\label{Eq:noisefloor}
k_\mathrm{CM} =\,(-0.9 \pm
6.2)\times\,10^{-21}\,\mathrm{T}^{-2},
\end{equation}
at 3$\sigma$ confidence level. As said before we give error bars at
$3\sigma$ corresponding to a confidence level of 99.8\%, that usually indicates an evidence for a non-zero signal. The uncertainty
takes into account the A-type and B-type uncertainties. The A-type
uncertainties come from the fit and from the photon lifetime with
a relative variation lower than 6$\%$ at $3\sigma$. The B-type
uncertainties have been evaluated previously and detailed in
Ref.\,\cite{Berceau2012}. They essentially come from the length of
the magnetic field $L_B$ with a relative uncertainty of 6.6$\%$ at
$3\sigma$. The value of Eq.\,(\ref{Eq:noisefloor}) gives an estimate of our noise
floor, which is half the one of the PVLAS collaboration in 2012
obtained with an integration time of 8192\,s \cite{Zavattini2012}.

In order to assess more precisely the physical origin of the
systematic effect, we zoom in the power spectral density of
$\Psi$, depicted in Fig.\,\ref{Fig:DSP_Psi}, on the frequencies
around 180\,Hz. We find several resonances at 177\,Hz, 200\,Hz and
above. The signal $J_4'$ is then fitted by a sine function but
with the frequency fixed to each of the resonance frequencies. The
best fit, corresponding to the best $\chi^2$, is obtained for
177\,Hz, which is compatible with the frequency given by the
previous fit. Fitting the residues by $\alpha B^2_\mathrm{f}$ gives our
final value for the CM constant:
\begin{eqnarray}
k_\mathrm{CM} = (5.1 \pm 6.2) \times
10^{-21}\,\mathrm{T}^{-2},\label{Eq:limits}
\end{eqnarray}
at 3$\sigma$ confidence level.

On the other hand, if we fit the data corresponding to Fig.\,\ref{Fig:J4p_fit_B2f} with the sum of the sine function of 177 Hz frequency and $\alpha B^2_\mathrm{f}$, we obtain:
\begin{eqnarray}
k_\mathrm{CM} = (8.3 \pm 8.0) \times
10^{-21}\,\mathrm{T}^{-2},\label{Eq:limits2}
\end{eqnarray}
at 3$\sigma$ confidence level.

All this shows that our noise floor given by the uncertainties is
of a few $10^{-21}\,\mathrm{T}^{-2}$ while the central value
depends on the fitting procedure. Establishing what is the most
statistically appropriate fitting procedure is out of the scope of
this paper. Our goal is to report on our noise floor and to
highlight the main contributions to systematic effects in order to
improve the overall sensitivity of the next version of the
apparatus.

\begin{figure}[t]
\begin{center}
\includegraphics[width=8cm]{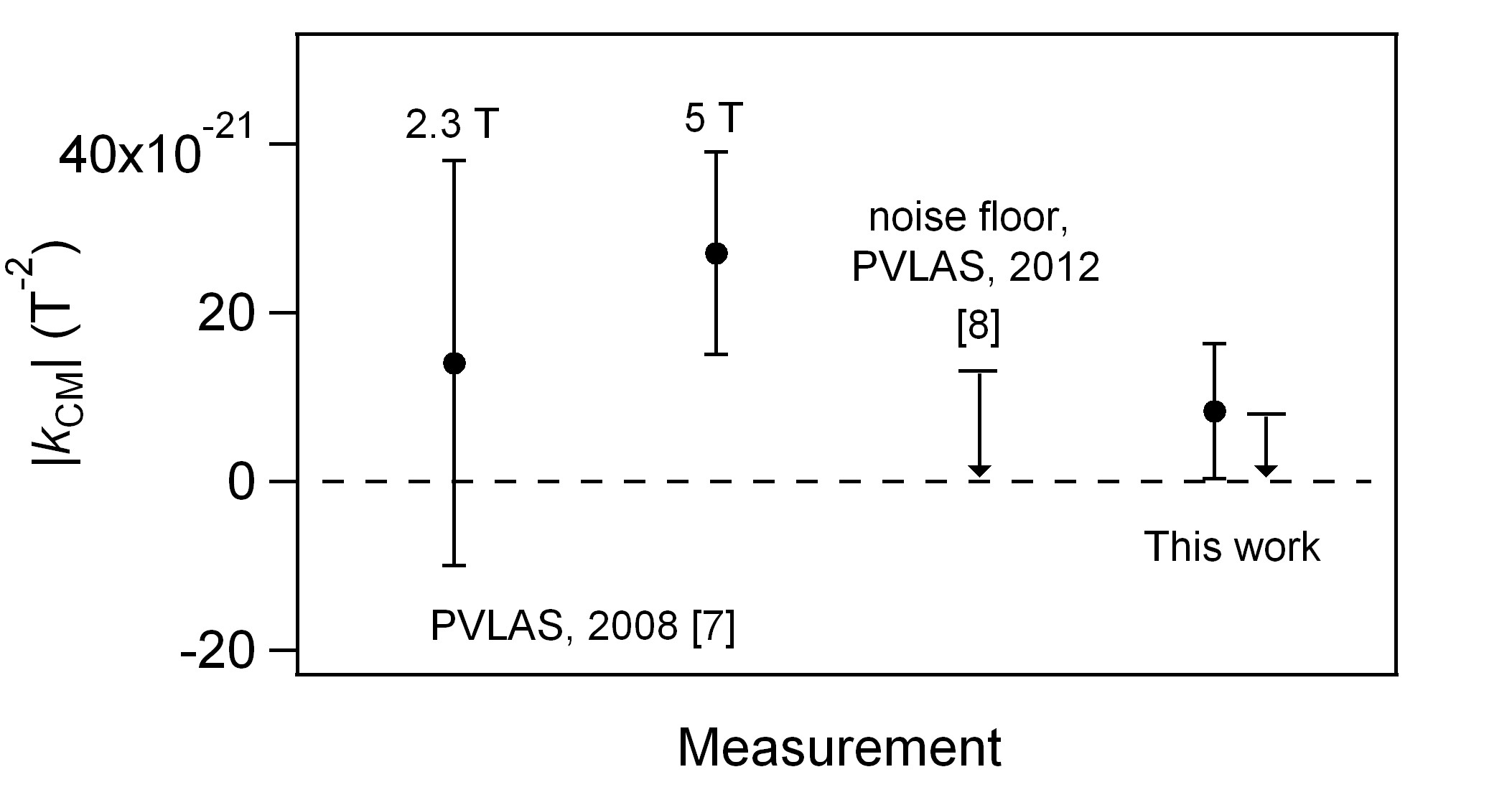}
\caption{\label{Fig:Comparison} Comparison of the latest absolute
reported values of the vacuum CM effect. Error bars are given at
3$\sigma$. This work: black dot, value obtain with the fit combining the sine function at 177 Hz and the $\alpha B^2_\mathrm{f}$ function; arrow, noise floor of $8.0 \times
10^{-21}\,\mathrm{T}^{-2}$.}
\end{center}
\end{figure}

Nevertheless, for the sake of comparison we show in
Fig.\,\ref{Fig:Comparison} our typical value given in Eq.\,(\ref{Eq:limits2}) together with the already published values. We see that our value is slightly better than the previous one.

\section{Conclusions and perspectives}

We presented the last advances of our BMV apparatus in terms of
the best noise floor of vacuum magnetic birefringence ever
realized. Our result validates our experimental method based on
pulsed fields. In particular, it proves that the sensitivity
obtained in a single pulse compensates the loss of duty cycle. To
reach the QED value, the needed improvement is of three orders of
magnitude. This is not conceivable with this first-generation
experiment. Our strategy is therefore to increase the magnetic
field thanks to the pulsed technology. At the moment, we have
$B^2L_B = 5.8$\,T$^2$m but we conceptualized and tested a pulsed
coil that has already reached a $B^2L_B$ higher than 300\,T$^2$m.
Two coils of this type will be inserted in the experiment in the
near future. This essential step really makes the vacuum
birefringence measurement within our reach.

On the other hand, our analysis has allowed us to identify some
systematic effects. Obviously, a special care will be devoted to
limit them in order to improve the accuracy. The magnetic field
induces an excitation on the setup which resonates at different
frequencies. Since it affects the signal $J_4$, the resonance at
177\,Hz has an odd symmetry with respect to the sign of $\Gamma$.
This implies that it concerns the mirror mounts. In order to get
rid of this effect, a new setup was designed, providing a better
magnetic insulation of the mirrors. It will also provide a better
acoustic insulation of the mirror mounts, improving the overall
sensitivity and decreasing the number of rejected shots. Moreover in the new version of our setup we will be able to measure the ellipticity both with $\theta_\mathrm{P}$ equal to $0^\circ$ (no induced ellipticity) and $45^\circ$ (maximal induced ellipticity). This will allow us to subtract from the raw data the systematic effects that do not depend on the polarization direction, as the sine function at 177 Hz.

\section{Acknowledgments}

We thank all the members of the BMV collaboration, and in
particular J. B\'eard, J. Billette, P. Frings, B. Griffe, J.
Mauchain, M. Nardone, J.-P. Nicolin and G. Rikken for strong
support. We are also indebted to the whole technical staff of
LNCMI. We acknowledge the support of the \textit{Fondation pour la
recherche IXCORE} and the ANR-Programme non Th\'ematique (Grant
No. ANR-BLAN06-3-139634).

\end{document}